\documentclass[a4paper]{article}
\usepackage{smsi}
\usepackage{comment}
\usepackage{graphicx}
\usepackage{amsmath}

\begin{document}

\title{Adaptive algorithm for microsensor in sustainable environmental monitoring}
\author{Nursultan Daupayev\textsuperscript{1},  Christian Engel\textsuperscript{2}, Ricky Bendyk\textsuperscript{2}, Prof. Dr.-Ing. Sören Hirsch\textsuperscript{1}}

\renewcommand{\institute}{\textsuperscript{1}  Brandenburg University of Applied Science, Department of Engineering, Magdeburger Str. 50, 14770
Brandenburg/Havel, Germany \\}

\renewcommand{\email}{Corresponding Author’s e-mail address: nursultan.daupayev@th-brandenburg.de}
          
\renewcommand{\abstractt}{
    Traditional data collection from sensors produce a lot of data, which lead to constant power consumption and require more storage space. This study proposes an algorithm for a data acquisition and processing method based on Fourier transform (DFT), which extracts dominant frequency components using harmonic analysis (HA) to identify frequency peaks. This algorithm allows sensors to activate only when an event occurs, while preserving critical information for detecting defects, such as those in the surface structures of buildings and ensuring accuracy for further predictions.
    }

\renewcommand{\keywords}{Microsensor, data reduction, anomaly detection, fourier transform, harmonic analysis}

\maketitle

\section{Introduction}

Continuous collection of environmental data in sensors plays an important role for example for anomaly detection in building structures [1]. However, regular data collection results in high energy consumption and data redundancy, where it is necessary to either optimize data in the hardware storage or delete them, which leads to a loss of data quality [2]. 


\section{Description of the method}

Within the scope of applying the algorithm, we collected different parameters including carbon dioxide ($CO_2$), humidity ($H$) and temperature ($T$) from a micro multivariate sensor, shown in Fig. 1, installed on the surface of the wall inside the building's room.

\begin{figure}[ht!]
    \centering
    \includegraphics[width=\linewidth]{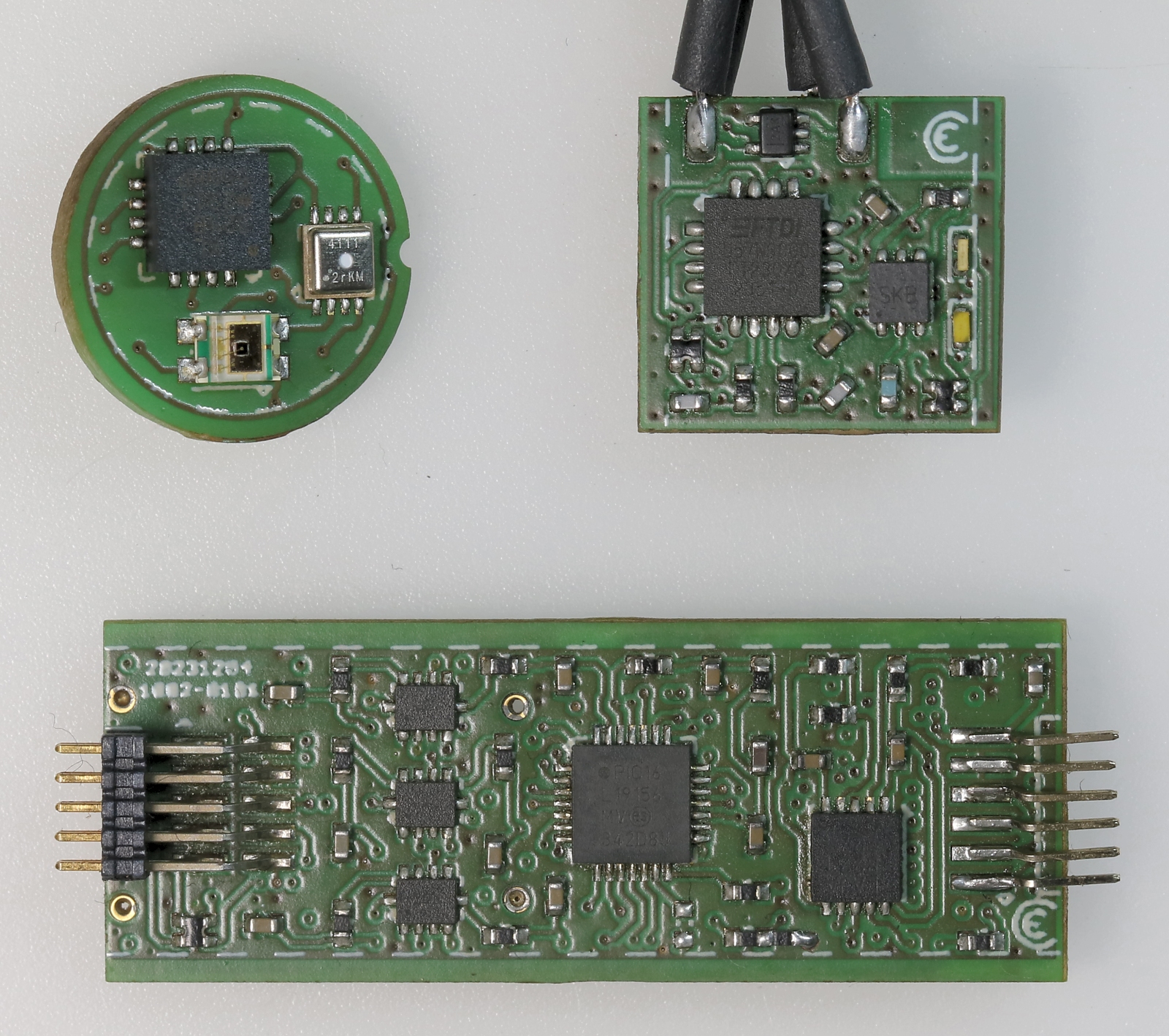}
    \caption{Multivariate Microsensor}
    \label{fig:ex1}
\end{figure}

\noindent
Initially, the algorithm analyses the data behaviour and it can be represented as a time series graph for $CO_2$, $T$, $H$ (Fig. 2) and simulation (Fig. 3) for $CO_2$ based on 24-hour measurements with an interval of 15 minutes. In this study, we focus only on the $CO_2$, since the application of the method on all parameters is similar.



\begin{figure}[ht!]
    \centering
    \includegraphics[width=\linewidth]{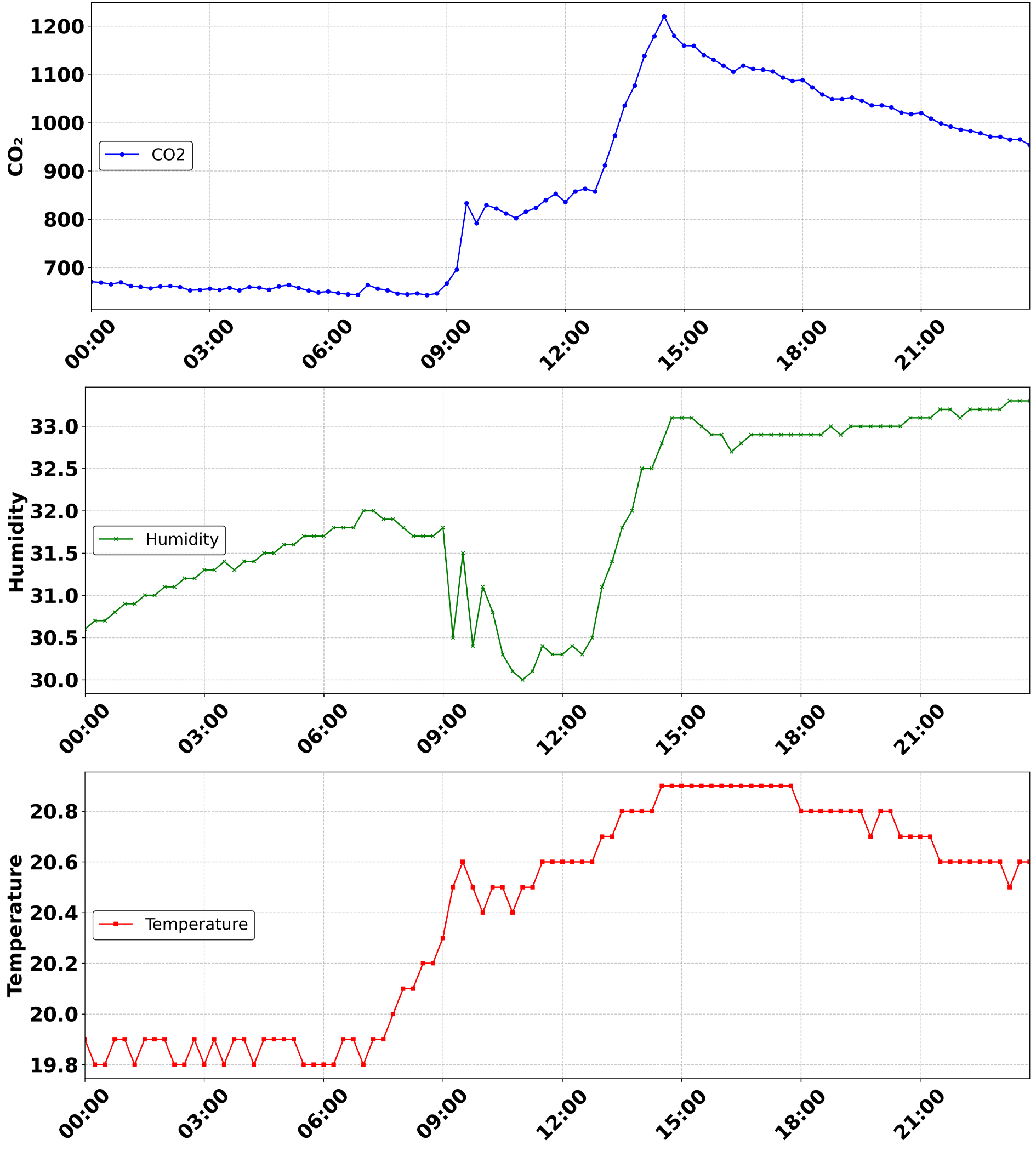}
    \caption{Microsensor Measurements}
    \label{fig:ex2}
\end{figure}

\begin{figure}[ht!]
    \centering
    \includegraphics[width=\linewidth]{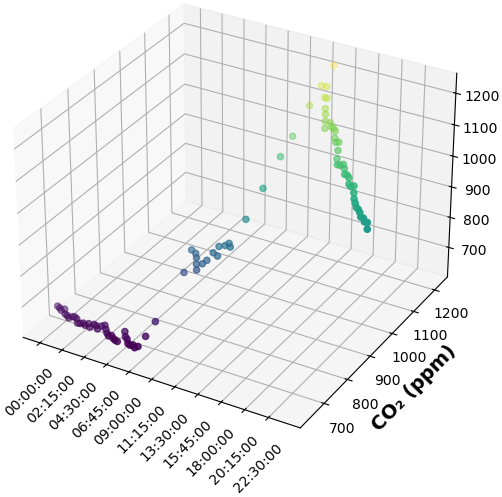}
    \caption{$CO_2$ Modeling}
    \label{fig:ex5}
\end{figure}

\noindent
Following this, the developed algorithm is aimed at reducing the number of measurements by finding important peaks and excluding insignificant measurement points. To achieve this, by applying  Fast Fourier Transform (FFT), algorithm transforms the data from the time domain to the frequency domain, which made it possible to identify the periodic characteristics of the sensor signals: 


\begin{equation}
X[k]=\sum_{n=0}^{N-1} x[n] e^{-j \frac{2 \pi}{N} k n}, \quad k=0,1, \ldots, N-1 .    
\end{equation}


\noindent
Then, it determines the optimal number of harmonics to achieve a reduction in sensor measurement, where the optimal number of harmonics is selected. In this case, there are 26 optimal harmonics:

\begin{equation}
A[k]=|X[k]|=\sqrt{(\operatorname{Re}\{X[k]\})^2+(\operatorname{Im}\{X[k]\})^2}
\end{equation}

\noindent
After selecting the optimal harmonics, an algorithm uses an inverse fast Fourier transform (IFFT) to reconstruct the signal (Fig 4.). It is expected to be close to the original data, showing minor deviations from them:

\begin{equation}
x[n]=\frac{1}{N} \sum_{k=0}^{N-1} X[k] e^{j \frac{2 \pi}{N} k n}
\end{equation}

\begin{figure}[ht!]
    \centering
    \includegraphics[width=\linewidth]{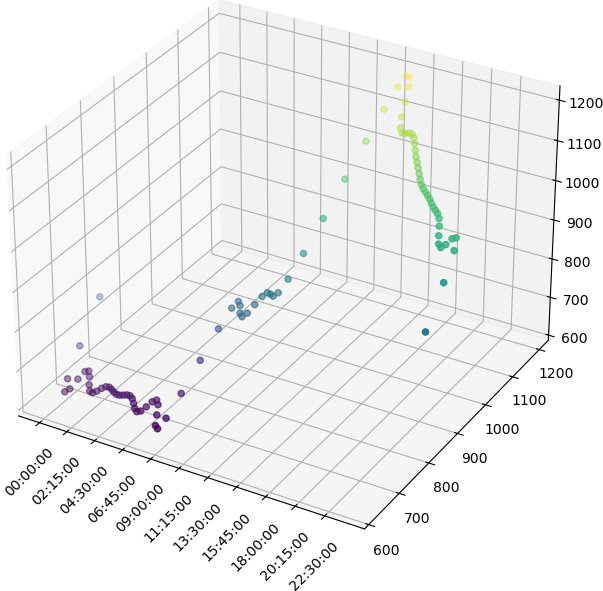}
    \caption{Reconstructed $CO_2$ Signal}
    \label{fig:ex4}
\end{figure}

\noindent
In addition, we are also aim to save energy by preserving at least 50\% of the total signal energy (in current data measurements), while minimizing the amount of data by eliminating unimportant harmonics and saving the main signal trends:

\begin{equation}
\sum_{k=0}^{K_{\text {opt }}}|X[k]|^2 \geq 0.5 \sum_{k=0}^{N-1}|X[k]|^2
\end{equation}

\section{Results}
As a result, the necessary points for sensor activation were obtained by applying the DFT and HA, that allowed to reduce the number of measurements, while saving important frequency characteristics of the signal (Fig.5).


\begin{figure}[ht!]
    \centering
    \includegraphics[width=\linewidth]{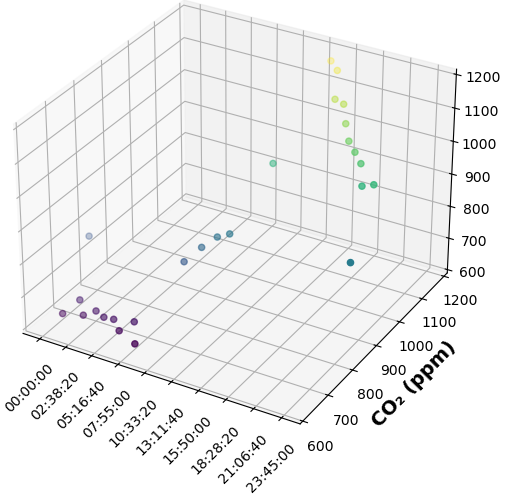}
    \caption{Microsensor activation time for $CO_2$}
    \label{fig:ex3}
\end{figure}

\end{document}